# Endohedral resonances: modification of atomic photoionization by the fullerenes shell

M. Ya. Amusia[1,2] and L. V. Chernysheva[2]

[1]Racah Institute of Physics, the Hebrew University, Jerusalem 91904, Israel
[2]Ioffe Physical-Technical Institute, St.-Petersburg 194021, Russia

E-mail: amusia@vms.huji.ac.il

**Abstract.** We discuss the complicated resonance structure of the endohedral atom photoionization cross section. Very strong enhancement and interference patterns in the photoionization cross-section of the valent and subvalent subshells of noble gas endohedral atoms A@$C_{60}$ are demonstrated. It is shown also that the atomic Giant resonance can be either completely destroyed or remains almost untouched depending on the velocity of photoelectrons that are emitted in the resonance's decay process.

These effects are results of dynamic modification of the incoming beam of radiation due to polarization of the fullerenes electron shell and reflection of photoelectrons be the fullerenes shell static potential.

We have considered the outer *np*- and subvalent *ns*-subshells for Ne, Ar, Kr and Xe noble gas atoms. The modification of the Giant resonances are considered for a whole sequence of endohedrals with atoms and ions Xe, Ba, La, $Ce^+$, $Ce^{+4}$, Eu.

The polarization of the fullerene shell is expressed via the total photoabsorption cross section. The photoelectron reflection from the static potential is taken into account in the frame of the so-called bubble potential that is a spherical $\delta$-type potential.

## 1. Introduction

Soon after discovery of fullerenes in 1985 it became clear that inside its empty shell an atom or even a small molecule can be "caged". Today inside a fullerene almost any element from the periodic table can be stuffed, thus forming new objects for scientific research and technological applications. From the point of view of photoionization studies they are similar to multi-shell heavy atoms. So, it is not surprising that a great deal of attention has been and still is concentrated on photoionization of endohedral atoms. It was demonstrated in a number of papers [1-6] that the $C_{60}$ shell adds prominent resonance structure in the photoionization cross section of endohedral atoms. However, since their size is considerably bigger than that of ordinary atoms, and fullerenes themselves consist of many carbon atoms, the direct and accurate quantum-mechanical calculations of an endohedrals are extremely difficult.

Therefore, a model approach is inevitable. As such, we consider, in at least qualitative agreement with the existing experimental data that for endohedral ionization by photons with energies from thresholds up to several hundreds electron-volts, one can substitute the real $C_{60}$ structure by a spherically - symmetric layer of approximately 240 electrons that move in the field of homogeneously smeared positive charge of carbon nuclei screened each by two 1s-electrons. This simplification permits to deal with endohedrals photoionization reasonable accurate thus believing that the results obtained are at least qualitatively reliable.

In this paper we will consider the complicated resonance structure of the endohedral atom photoionization cross section. Namely, very strong enhancement and interference patterns in the photoionization cross-section of the valent and subvalent subshells of noble gas endohedral atoms A@$C_{60}$ are demonstrated. It is shown also that the atomic Giant resonance can be either completely destroyed or remains almost untouched depending on the velocity of photoelectrons that are emitted in the resonance's decay process.

It exists a deep similarity between a multi-electron atom and an endohedral from the point of view of the multi-electron effects. Using the presented above simplified description of the electron fullerenes shell, an endohedral can be treated as an atom with an additional multi-electron shell. In the same way one can consider multi-layer fullerenes or so-called fullerenes onions, in which, in the central empty volume the caged atom is located.

The pronounced action of the multi-electron neighboring shell upon a few-electron one was considered for the first time thirty-five years ago. As an example, the influence of $3p^6$ electrons upon the $3s^2$ in Ar has been presented [7]. A more complicated case with three interacting subshells was considered in [8]. It was demonstrated that the $5p^6$ and $4d^{10}$ subshells act upon $5s^2$ in Xe very strong, completely modifying the $5s^2$ photoionization cross-section. All corresponding calculations were performed in the frame of the so-called Random Phase Approximation with Exchange (RPAE). The first experimental confirmations of these predictions were obtained soon [9]. Since then the investigations of the effects of inter-shell interaction in atoms have become a permanent subject of research (see, e.g. [10, 11]).

The physical nature of these inter-shell effects in photoionization is as follows. A many-electron atomic subshell is polarized by an electromagnetic wave and a dipole moment is induced in it. Under the action of this dipole moment a neighboring atomic subshell is ionized. RPAE is extremely convenient to describe this effect. So, the ionization of a given electron can proceed via several pathways: directly, after photon absorption by the ionizing electron, and indirectly, in two or even several steps, via virtual excitation of other subshells. Since the electronic subshells in an atom are not separated spatially well enough, the amplitude of these two- or multi-step photo-processes cannot be expressed accurately enough via the dipole polarizability of the many-electron subshells.

In this sense the situation for the endohedral atoms A@$C_n$ is quite different. Of course, it consists of very many electrons, to large extent delocalized, and nuclei, thus being a hard object for *ab initio* treatment as compared to an isolated atom. However, in endohedrals an essential simplifying factor exists. Namely, the radius of the fullerene shell significantly exceeds that of an encapsulated atom. This makes it possible for photoionization of the A atom, in the first approximation, to consider the electronic sub-systems of the fullerene shell (or several shells in fullerene onions) and atom as practically independent of each other. For this reason, the amplitude of atom photoionization going through virtual excitation of fullerene shell electrons can be expressed directly via the dynamic polarizability of the fullerene shell $\alpha_{C_{60}}^d(\omega)$, or in case of "onions". In those cases when the frequency of electromagnetic radiation is close to frequencies of plasma oscillations of the collectivized electrons of the fullerene, the role of this two-step process becomes decisively important, as the role of $4d^{10}$ upon $5s^2$ in isolated Xe.

Along with fullerenes shell polarization, one has to take into account also the reflection and refraction of the photoelectron wave, which goes from $np^6$ and $ns^2$ subvalent shell, by the static potential of the fullerene. This reflection leads to formation of oscillating pattern of the cross section (see e.g. [12, 13]).

The molecule A@$C_n$ is a remarkable concrete example illustrating the role of the inter-shell interactions in the fullerene-like molecules, qualitatively similar but even much stronger than in the isolated atoms. In this case three types of resonances, that in photoelectrons' reflection from the $C_{60}$ shell, plasmon-type excitation of generalized $C_{60}$ electrons and that caused by action of caged atoms neighboring to $np$ and $ns$-subshells, interfere.

We will discuss here three types of manifestation of the $C_{60}$ effect upon the endohedral atom photoionization. We will consider the action of the $C_{60}$ polarization by the incoming photon beam and photoelectron's reflection by the $C_{60}$ static potential upon photoionization of outer $np^6$ subshell of noble gases. We will show that the combination of both effects leads to increase of the cross-sections by a factor 20-30, creating powerful resonances that we call *Giant endohedral resonances*.

The effect of $C_{60}$ polarization on $ns^2$ is much less pronounced then on $np^6$ but still, together with static potential reflection adds very interesting structure that include also the effect of the neighboring $np^6$ subshell. A simple method was developed to take into account the reflection of photoelectron wave by $C_{60}$ static potential that was presented by of a zero-thickness $\delta$-type so-called bubble potential (see [16] and references there in). This approximation is valid for slow photoelectrons, whose wavelength is bigger than the thickness of $C_{60}$ shell.

Then we will consider the role $C_{60}$ upon the atomic Giant resonances. Since their energy is high enough, the role of $C_{60}$ polarization is negligible. Starting with Xe@$C_{60}$, where the Giant resonance is almost completely destroyed by the photoelectron's reflection, we will end up with Eu@$C_{60}$, where the atomic Giant resonance remains untouched.

Entirely, we will show in this paper that the dynamic polarization of $C_{60}$ drastically modifies the outer [17, 18] and subvalent [19] shell photoionization cross section at any frequency of the incoming radiation $\omega$. The photoionization cross-section of outer and subvalent shells of noble gas endohedral atoms drastically differ from respective data for isolated atoms. We will also trace the modification of the atomic Giant resonances under the $C_{60}$ shell action on the way from Xe@$C_{60}$ to Eu@ $C_{60}$ [20].

It would be of interest to see the alteration of the photoionization cross-section if instead of $C_{60}$ other fullerenes, like $C_{70}$, $C_{76}$, $C_{82}$ or $C_{87}$, are considered. However, to study the endohedrals with $C_{70}$, $C_{76}$, $C_{82}$ or $C_{87}$ one needs to know the shape of these objects, their photoionization cross sections and the location of the caged atoms inside fullerene. The answers to these questions are absent at this moment.

It is assumed in our following derivations that an atom is centrally located in the fullerene and in accordance with the available experimental data, that the fullerene radius is much bigger than the atomic radius and the thickness of the fullerene shell. As we will see, these assumptions permit for the A@$C_{60}$ photoionization cross section to be presented as a product of the atomic photoionization cross-section and two calculated factors that account for polarization of the fullerenes electron shell and reflection of photoelectrons by the fullerene static potential.

## 2. Essential formulae

We will use here the theoretical approaches already developed in a number of previous papers [12, 13]. However, for completeness, let us repeat the main points of the consideration and present the essential formula used in calculations.

Let us start with the problem of an isolated closed shell atom. For the differential in angle photoionization cross-section by linearly polarized light of frequency $\omega$ the following expressions [21] can be derived from the more general one [22]:

$$\frac{d\sigma_{nl}(\omega)}{d\Omega} = \frac{\sigma_{nl}(\omega)}{4\pi}[1 + \beta_{nl} P_2(\cos\Theta) + (\delta_{nl} + \gamma_{nl}\cos^2\Theta)\sin\Theta\cos\Phi]. \tag{1}$$

Here $\sigma_{nl}(\omega)$ is the $nl$-subshell partial cross-section, $P_2(\cos\Theta)$ is the Legendre polynomial, $\beta_{nl}(\omega)$ is the dipole, while $\gamma_{nl}(\omega)$ and $\delta_{nl}(\omega)$ are so-called non-dipole angular anisotropy parameters, $\Theta$ is the polar angle between the vectors of photoelectron's velocity **v** and photon's polarization **e**, while $\Phi$ is the azimuth angle determined by the projection of **v** in the plane orthogonal to **e** that includes the vector of photon's velocity.

There are two possible dipole transitions from subshell $l$, namely $l \to l\pm 1$ and three quadrupole transitions $l \to l; l\pm 2$. Corresponding general expressions for $\beta_{nl}(\omega)$, $\gamma_{nl}(\omega)$ and $\delta_{nl}(\omega)$ are rather complex and are presented as combinations of dipole $d_{l\pm 1}$ and quadrupole $q_{l\pm 2,0}$ matrix elements of photoelectron transitions and photoelectrons waves phases. In one-electron Hartree-Fock (HF) approximation these parameters are presented as [23]:

$$\beta_{nl}(\omega) = \frac{1}{(2l+1)\left[(l+1)d_{l+1}^2 + ld_{l-1}^2\right]}[(l+1)(l+2)d_{l+1}^2 + l(l-1)d_{l-1}^2 - $$
$$6l(l+1)d_{l+1}d_{l-1}\cos(\delta_{l+1} - \delta_{l-1})]. \quad (2)$$

Parameter $\gamma_{nl}(\omega)$ is given by the following expression

$$\gamma_{nl}(\omega) = -\frac{3\kappa}{\left[ld_{l-1}^2 + (l+1)d_{l+1}^2\right]}\left\{\frac{(l+1)(l+2)}{(2l+1)(2l+3)}q_{l+2}[5ld_{l-1}\cos(\delta_{l+2} - \delta_{l-1}) - \right.$$
$$-(l+3)d_{l+1}\cos(\delta_{l+2} - \delta_{l-1})] - \frac{(l-1)l}{(2l+1)(2l+1)}q_{l-2} \times$$
$$\times [5(l+1)d_{l+1}\cos(\delta_{l-2} - \delta_{l+1}) - (l-2)d_{l-1}\cos(\delta_{l-2} - \delta_{l-1})] + \quad (3)$$
$$+ 2\frac{l(l+1)}{(2l-1)(2l+3)}q_l[(l+2)d_{l+1}\cos(\delta_l - \delta_{l+1}) - (l-1)d_{l-1}\cos(\delta_l - \delta_{l-1})]\right\}.$$

As is shown by calculations, usually $\delta_{nl} \ll \gamma_{nl}$ and therefore we are not presenting the corresponding rather complex expression for $\delta_{nl}$.

In (3) $\kappa = \omega/c$, $\delta_l(k)$ are the photoelectrons' scattering phases; the following relation gives the matrix elements $d_{l\pm 1\uparrow\downarrow}$ in the so-called $r$-form

$$d_{l\pm 1} \equiv \int_0^\infty P_{nl}(r) r P_{\varepsilon l\pm 1}(r) dr, \quad (4)$$

where $P_{nl}(r)$, $P_{\varepsilon l\pm 1}(r)$ are the radial Hartree-Fock (HF) [23] one-electron wave functions of the $nl$ discrete level and $\varepsilon l\pm 1$ - in continuous spectrum, respectively. The following relation gives the quadrupole matrix elements

$$q_{l\pm 2,0} \equiv \frac{1}{2}\int_0^\infty P_{nl}(r) r^2 P_{\varepsilon l\pm 2,0}(r) dr. \quad (5)$$

In order to take into account the Random Phase Approximation with Exchange (RPAE) [23] multi-electron correlations, one has to perform the following substitutions in the expressions for $\beta_{nl}(\omega)$ and $\gamma_{nl}(\omega)$ [24]:

$$d_{l+1}d_{l-1}\cos(\delta_{l+1}-\delta_{l-1}) \rightarrow [(\operatorname{Re}D_{l+1}\operatorname{Re}D_{l-1}+\operatorname{Im}D_{l+1}\operatorname{Im}D_{l-1})\cos(\delta_{l+1}-\delta_{l-1})-$$
$$-(\operatorname{Re}D_{l+1}\operatorname{Im}D_{l-1}-\operatorname{Im}D_{l+1}\operatorname{Re}D_{l-1})\sin(\delta_{l+1}-\delta_{l-1})] \equiv \quad (6)$$
$$\equiv \tilde{D}_{l+1}\tilde{D}_{l-1}\cos(\delta_{l+1}+\Delta_{l+1}-\delta_{l-1}-\Delta_{l-1}).$$

$$d_{l\pm 1}q_{l\pm 2,0}\cos(\delta_{l\pm 2,0}-\delta_{l\pm 1}) \rightarrow [(\operatorname{Re}D_{l\pm 1}\operatorname{Re}Q_{l\pm 2,0}+\operatorname{Im}D_{l\pm 1}\operatorname{Im}Q_{l\pm 2,0})\cos(\delta_{l\pm 2,0}-\delta_{l\pm 1})-$$
$$-(\operatorname{Re}D_{l\pm 1}\operatorname{Im}Q_{l\pm 2,0}-\operatorname{Im}D_{l\pm 1}\operatorname{Re}Q_{l\pm 2,0})\sin(\delta_{l\pm 2,0}-\delta_{l\pm 1})] \equiv$$
$$\equiv \tilde{D}_{l\pm 1}\tilde{Q}_{l\pm 2,0}\cos(\delta_{l\pm 2,0}+\Delta_{l\pm 2,0}-\delta_{l\pm 1}-\Delta_{l\pm 1}), \quad (7)$$
$$d_{l\pm 1}^2 \rightarrow \operatorname{Re}D_{l\pm 1}^2+\operatorname{Im}D_{l\pm 1}^2 \equiv \tilde{D}_{l\pm 1}^2.$$

Here the following notations are used for the matrix elements with account of multi-electron correlations, dipole and quadrupole, respectively:

$$D_{l\pm 1}(\omega) \equiv \tilde{D}_{l\pm 1}(\omega)\exp[i\Delta_{l\pm 1}(\varepsilon)], \quad Q_{l\pm 2,0}(\omega) \equiv \tilde{Q}_{l\pm 2,0}(\omega)\exp[i\Delta_{l\pm 2,0}(\varepsilon)], \quad (8)$$

where $\tilde{D}_{l\pm 1}(\omega)$, $\tilde{Q}_{l\pm 2,0}(\omega)$, $\Delta_{l\pm 1}$ and $\Delta_{l\pm 2,0}$ are absolute values of the amplitudes for respective transitions and phases for photoelectrons with angular moments $l\pm 1$ and $l\pm 2,0$.

The following are the RPAE equation for the dipole matrix elements [23]

$$\langle v_2|D(\omega)|v_1\rangle = \langle v_2|d|v_1\rangle + \sum_{v_3,v_4} \frac{\langle v_3|D(\omega)|v_4\rangle(n_{v_4}-n_{v_3})\langle v_4 v_2|U|v_3 v_1\rangle}{\varepsilon_{v_4}-\varepsilon_{v_3}+\omega+i\eta(1-2n_{v_3})}, \quad (9)$$

where

$$\langle v_1 v_2|U|v_1' v_2'\rangle \equiv \langle v_1 v_2|V|v_1' v_2'\rangle - \langle v_1 v_2|V|v_2' v_1'\rangle. \quad (10)$$

Here $V \equiv 1/|\vec{r}-\vec{r}'|$ and $v_i$ is the total set of quantum numbers that characterize a HF one-electron state on discrete (continuum) levels, $\varepsilon_{v_i}$ is the HF energy, $\eta \rightarrow +0$. That includes the principal quantum number (energy), angular momentum, its projection and the projection of the electron spin. The function $n_{v_i}$ (the so-called step-function) is equal to 1 for occupied and 0 for vacant states.

The dipole matrix elements $D_{l\pm 1}$ are obtained by solving the radial part of the RPAE equation (12). As to the quadrupole matrix elements $Q_{l\pm 2,0}$, they are obtained by solving the radial part of the RPAE equation, similar to (12)

$$\langle v_2|Q(\omega)|v_1\rangle = \langle v_2|\hat{q}|v_1\rangle + \sum_{v_3,v_4} \frac{\langle v_3|Q(\omega)|v_4\rangle(n_{v_4}-n_{v_3})\langle v_4 v_2|U|v_3 v_1\rangle}{\varepsilon_{v_4}-\varepsilon_{v_3}+\omega+i\eta(1-2n_{v_3})}. \quad (11)$$

Here in r-form one has $\hat{q} = r^2 P_2(\cos\theta)$.
Equations (9, 11) are solved numerically using the procedure discussed at length in [25].

### 3. Effect of C$_{60}$ fullerene shell

Let us start with the confinement effects. These effects near the photoionization threshold can be described within the framework of the "orange" skin potential model. According to this model, for small photoelectron energies the real static and not perfectly spherical potential of the $C_{60}$ can be presented by the zero-thickness bubble pseudo-potential (see [26] and references therein):

$$V(r) = -V_0 \delta(r - R). \tag{12}$$

The parameter $V_0$ is determined by the requirement that the binding energy of the extra electron in the negative ion $C_{60}^-$ is equal to its observable value. Addition of the potential (12) to the atomic HF potential leads to a factor $F_l(k)$ in the photoionization amplitudes, which depends only upon the photoelectron's momentum $k$ and orbital quantum number $l$ [26]:

$$F_l(k) = \cos \breve{\Delta}_l(k) \left[ 1 - \tan \breve{\Delta}_l(k) \frac{v_{kl}(R)}{u_{kl}(R)} \right], \tag{13}$$

where $\breve{\Delta}_l(k)$ are the additional phase shifts due to the fullerene shell potential (12). They are expressed by the following formula:

$$\tan \breve{\Delta}_l(k) = \frac{u_{kl}^2(R)}{u_{kl}(R) v_{kl}(R) + k/2V_0}. \tag{14}$$

In these formulas $u_{kl}(r)$ and $v_{kl}(r)$ are the regular and irregular solutions of the atomic HF equations for a photoelectron with momentum $k = \sqrt{2\varepsilon}$, where $\varepsilon$ is the photoelectron energy connected with the photon energy $\omega$ by the relation $\varepsilon = \omega - I_A$ with $I_A$ being the atom A ionization potential.

Using Eq. (13), one can obtain the following relation for $D^{AC(r)}$ and $Q^{AC(r)}$ amplitudes for endohedral atom A@$C_{60}$ with account of photoelectron's reflection and refraction by the $C_{60}$ static potential (12), expressed via the respective values for isolated atom that correspond to $nl \to \varepsilon l'$ transitions:

$$\begin{aligned} D_{nl,kl'}^{AC(r)}(\omega) &= F_{l'}(k) D_{nl,kl'}(\omega), \\ Q_{nl,kl''}^{AC(r)}(\omega) &= F_{l''}(k) Q_{nl,kl''}(\omega) \end{aligned}. \tag{15}$$

For the cross-sections one has

$$\sigma_{nl,kl'}^{AC(r)}(\omega) = [F_{l'}(k)]^2 \sigma_{nl,kl'}^A(\omega), \tag{16}$$

where $\sigma_{nl,kl'}^A(\omega)$ is the contribution of the $nl \to \varepsilon l'$ transition to the photoionization cross-section of atomic subshell $nl$, $\sigma_{nl}^A(\omega)$.

Now let us discuss the role of polarization of the $C_{60}$ shell under the action of the photon beam [15]. The effect of the fullerene electron shell polarization upon atomic photoionization amplitude can be taken into account in RPAE using (9). This approximation is good for isolated atoms [23] and it is reasonable to assume that it is good also for endohedral atoms as well.

Symbolically, the total amplitude of electron photoionization of the "caged" atom $D_A$ can be presented by applying (9) to the whole endohedral system as a sum of two terms

$$\hat{D}_A = \hat{d}_A + \hat{D}_C \hat{\chi} U_{CA}, \qquad (17)$$

where $\hat{D}_C$ is the ionization amplitude of any other than "A"-electrons, $\hat{\chi} = 1/(\omega - \hat{H}_{ev}) - 1/(\omega + \hat{H}_{ev})$ is the propagator of other electron excitation, i.e. electron ($e$)-vacancy ($v$) pair creation, $\hat{H}_{ev}$ is the pair Hartree-Fock Hamiltonian. The interaction (10) can be presented as $U_{CA} \equiv V_{CAdir} - V_{CAexc}$, with $V_{CAdir}$ and $V_{CAexc}$ being the operator of direct and exchange pure Coulomb interaction between "C" and "A" electrons.

The formula (17) is simplified considerably if the "A"-electrons are at much smaller distances from the centre of the system than the "C"-electrons. Then the Coulomb interaction is considerably simplified, becoming

$$U_{CA} \approx \mathbf{r}_C \cdot \mathbf{r}_A / r_C^3, \; (r_C \gg r_A). \qquad (18)$$

Here $\mathbf{r}_A$ and $\mathbf{r}_C$ are the "A"- and "C"-electron shells radii, respectively.

The effect of the "C"-shell is represented particularly simple, when it is an outer one, located well outside the intermediate and outer atomic subshell. Then rightfully neglecting the exchange "A-C"-interaction and representing $U_{AC}$ as (18), one reduces (17) to an algebraic equation instead of operator one where $\hat{D}_C \hat{\chi} U_{CA}$ is substituted by the following expression

$$[2 \sum_{evexit,C} \omega_{ev} D_{ev}(\omega)(\omega^2 - \omega_{ev}^2)^{-1} d_{ev}] / \bar{r}_C^3 \equiv -\alpha_C(\omega) / \bar{r}_C^3. \qquad (19)$$

Here the summation over $evexit,C$ goes over all electron-vacancy excitation of the considered shell. Some more complex excitations are included into the amplitude $D_{ev}(\omega)$ and $\bar{r}_C$ is the mean radius of the "C"-shell that coincide with the fullerene radius $R_C$, the C$_{60}$ shell is thin enough.

Thus, one has instead of Eq. (17) the following formula [15]:

$$D_A(\omega) \cong d_A \left(1 - \frac{\alpha_C(\omega)}{R_C^3}\right), \qquad (20)$$

The amplitude of endohedral atom's photoionization due to $nl \to \varepsilon l'$ transition $D_{nl,\varepsilon l'}^{A@C}(\omega)$ with all essential atomic correlations taken into account can be presented by the following formula [15]:

$$D_{nl,\varepsilon l'}^{A@C}(\omega) \cong F_{l'}(k)\left(1 - \frac{\alpha_C^d(\omega)}{R_C^3}\right) D_{nl,\varepsilon l'}^A(\omega) \equiv F_{l'}(k) G^d(\omega) D_{nl,\varepsilon l'}^A(\omega), \qquad (21)$$

where $\alpha_C^d(\omega)$ is the dipole dynamical polarizability of C$_{60}$ and $R_C$ is its fullerenes radius and $D_{nl,\varepsilon l'}^A(\omega)$ accounts for all the electron correlations in the "caged" atom. For the quadrupole amplitude one obtains, starting from (11), a similar expression:

$$Q_{nl,\varepsilon l'}^{A@C}(\omega) \cong F_{l'}(k)\left(1 - \frac{\alpha_C^q(\omega)}{R_C^5}\right) Q_{nl,\varepsilon l'}^A(\omega) \equiv F_{l'}(k) G^q(\omega) Q_{nl,\varepsilon l'}^A(\omega), \qquad (22)$$

where $\alpha_C^q(\omega)$ is the quadrupole dynamical polarizability of C$_{60}$. The $G^{d,q}(\omega)$ factors are complex numbers that we present as

$$G^{d,q}(\omega) = \tilde{G}^{d,q}(\omega) \exp[i\eta^{d,q}(\omega)], \qquad (23)$$

where $\tilde{G}^{d,q}(\omega)$ are respective absolute values.

Using the relation between the imaginary part of the polarizability and the dipole photoabsorption cross-section $\sigma_C^d(\omega)$ - $\mathrm{Im}\,\alpha_C^d(\omega) = c\sigma_C^d(\omega)/4\pi\omega$, one can derive the polarizability of the C$_{60}$ shell. Although experiments [27] do not provide absolute values of $\sigma_C^d(\omega)$, it can be reliably estimated using different normalization procedures on the basis of the sun rule: $(c/2\pi^2)\int_{I_C}^{\infty}\sigma_C^d(\omega)d\omega = N$, where $N$ is the number of collectivized electrons, 240 for C$_{60}$. . The real part of polarizability is connected with imaginary one (and with the photoabsorption cross-section) by the dispersion relation:

$$\mathrm{Re}\,\alpha_C^d(\omega) = \frac{c}{2\pi^2}\int_{I_C}^{\infty}\frac{\sigma_C^d(\omega')d\omega'}{\omega'^2 - \omega^2}, \qquad (24)$$

where $I_C$ is the C$_{60}$ ionization potential.

The equality $\mathrm{Im}\,\alpha_C^q(\omega) = c\sigma_C^q(\omega)/4\pi\omega$ and quadrupole dispersion relation similar to (24) are valid. But the quadrupole photoabsorption cross-section is so small that it cannot be derived experimentally.

Note that because we assume the strong inequality $R_C \gg r_A$ ($r_A$ being the atomic radius) we have derived formulas (17) and (18) that are more accurate than those obtained from the RPAE for the whole A@F system. This is important since "one electron – one vacancy" channel that is the only taken into account in RPAE is not always dominant in the photoabsorption cross-section of the fullerene and hence in its polarizability.

Using the amplitude (26), one has for the cross section

$$\sigma_{nl,\varepsilon l'}^{A@C}(\omega) = [F_{l'}(\omega)]^2 \left|1 - \frac{\alpha_C^d(\omega)}{R_C^3}\right|^2 \sigma_{nl,\varepsilon l'}^A(\omega) \equiv [F_{l'}(\omega)]^2 S(\omega)\sigma_{nl,\varepsilon l'}^A(\omega), \qquad (25)$$

where $S(\omega) = [\tilde{G}^d(\omega)]^2$ cab be called radiation enhancement parameter.

With these amplitudes, using the expressions (4-6) and performing the substitution (9, 10) we obtain the cross-sections for NG@C$_{60}$ and angular anisotropy parameters. While calculating the anisotropy parameter, the cosines of atomic phases differences $\cos(\delta_l - \delta_{l'})$ in formulas (4)-(6) are replaced at first by $\cos(\delta_l + \Delta_l - \delta_{l'} - \Delta_{l'})$. As a result, one has for the dipole angular anisotropy parameter (4), using (9):

$$\beta_{nl}(\omega) = \frac{1}{(2l+1)\left[(l+1)F_{l+1}^2 \tilde{D}_{l+1}^2 + lF_{l-1}^2 \tilde{D}_{l-1}^2\right]}[(l+1)(l+2)F_{l+1}^2 \tilde{D}_{l+1}^2$$
$$+ l(l-1)F_{l-1}^2 \tilde{D}_{l-1}^2 - 6l(l+1)F_{l+1}F_{l-1}\tilde{D}_{l+1}\tilde{D}_{l-1}\cos(\tilde{\delta}_{l+1} - \tilde{\delta}_{l-1})] \quad . \quad (26)$$

where $\tilde{\delta}_{l'} = \delta_{l'} + \Delta_{l'}$ (see (11)). Naturally, the dipole parameter $\beta_{nl}(\omega)$ is not affected by $G^d(\omega)$ factors that similarly alter the nominator and denominator in (26).

The situation for non-dipole parameters is different, since $G^d(\omega) \neq G^q(\omega)$. From (5) and (6), using (9) and (10) we arrive to the following expressions for the non-dipole angular anisotropy parameters:

$$\eta_{nl}(\omega) = -\frac{3\kappa \tilde{G}^q(\omega)}{\tilde{G}^d(\omega)\left[(l+1)F_{l+1}^2 \tilde{D}_{l+1}^2 + lF_{l-1}^2 \tilde{D}_{l-1}^2\right]} \times$$
$$\times \left\{ \frac{(l+1)(l+2)}{(2l+1)(2l+3)} F_{l+2}\tilde{Q}_{l+2}\left[5lF_{l-1}\tilde{D}_{l-1}d_{l-1}\cos\left(\tilde{\tilde{\delta}}_{l+2} - \tilde{\tilde{\delta}}_{l-1}\right) - (l+3)F_{l+1}\tilde{D}_{l+1}\cos\left(\tilde{\tilde{\delta}}_{l+2} - \tilde{\tilde{\delta}}_{l-1}\right)\right] -$$
$$- \frac{(l-1)l}{(2l+1)(2l+1)} F_{l-2}\tilde{Q}_{l-2}\left[5(l+1)F_{l+1}\tilde{D}_{l+1}\cos\left(\tilde{\tilde{\delta}}_{l-2} - \tilde{\tilde{\delta}}_{l+1}\right) - (l-2)\tilde{F}_{l-1}\tilde{D}_{l-1}\cos\left(\tilde{\tilde{\delta}}_{l-2} - \tilde{\tilde{\delta}}_{l-1}\right)\right] +$$
$$+ 2\frac{l(l+1)F_l\tilde{Q}_l}{(2l-1)(2l+3)}\left[(l+2)F_{l+1}\tilde{D}_{l+1}\tilde{D}_{l+1}\cos\left(\tilde{\tilde{\delta}}_l - \tilde{\tilde{\delta}}_{l+1}\right) - (l-1)F_{l-1}\tilde{D}_{l-1}\tilde{D}_{l-1}\cos\left(\tilde{\tilde{\delta}}_l - \tilde{\tilde{\delta}}_{l-1}\right)\right]\right\},$$
$$(27)$$

where $\tilde{\tilde{\delta}}_{l\pm 2,0} = \tilde{\delta}_{l\pm 2,0} + \eta^q$ and $\tilde{\tilde{\delta}}_{l\pm 1} = \tilde{\delta}_{l\pm 1} + \eta^d$ (see (23)).

## 3. Results of calculations

The $C_{60}$ parameters in the present calculations were chosen the same as in the previous papers, e.g. in [15]: $R = 6.639$ and $V_0 = 0.443$. For this radius and the thickness of $C_{60}$ shell equal to 3, our approach is well justified for photoelectron energies of about 2-3 atomic units. However, for completeness and understanding the tendencies, we present data for higher energies also.

In Fig. 1 we present the radiation enhancement parameter $S(\omega)$ (see (12)), its amplitude's absolute value $\tilde{G}^d(\omega) \equiv |G(\omega)|$ and phase $\eta^d \equiv \arg G(\omega)$. The $np$ and $ns$ thresholds for Ne, Kr and Xe are located differently, strongly increasing all but the Ne 2s cross-sections

In all considered cases the influence of the fullerenes shell upon the photoionization of the "caged" atoms is prominent. Rather impressive are the oscillations due to reflection of the photoelectron by the fullerenes shell, described by the factor $F_{l'}(\omega)$. The influence of $S(\omega)$ is also big enough.

Although we performed calculations for all but He noble gases, in Fig. 2-5 we depict representative data on Ar 3p, 3s and Xe 3p, 3s data. We see from Fig. 2 and 3 that powerful resonance structures appeared in the outer shell photoionization cross-sections that we call Giant Endohedral resonances. It is remarkable that the cross-section reaches values up to about 1000 Mb that is by a factor 20-30 more than for isolated atoms and the corresponding sum rule of the resonance region is about 25, thus exciding the atomic value (6 for $np$-subshell) by a factor of 4. The influence of the $C_{60}$ shell upon angular anisotropy parameters is relatively weaker but oscillations due to photoelectron reflection are quite noticeable.

The effect of $S(\omega)$ and photoelectron's reflection upon subvalent subshells is illustrated in Fig. 4 and 5. Although not that big as on $np$, the modification of $ns$ photoionization parameters is quite prominent.

It was predicted in [28] and illustrated by Fig. 6 that the C60 shell due to reflection of photoelectrons destroys the 4d atomic Giant resonance. At respective photon energies the role of $S(\omega)$, as is seen from Fig. 1 is negligible. This prediction was checked in [29] but instead of Xe@$C_{60}$ the molecule Ce@$C_{82}$ was chosen. No structure, similar to given in Fig. 6 was found. We explain this by the fact that in Ce (in fact, according to [29], it is stripped to $Ce^{3+}$ inside $C_{60}$) the photoelectrons are fast, coming from the decay of 4$d$-4$f$ discrete excitation into outer shell continuous spectrum. This hypothesis was checked by calculating the photoionization cross-section for a sequence of endohedrals – from Xe@$C_{60}$ to Eu@$C_{60}$. The results are depicted in Fig. 6-9, confirming the hypothesis: in La@$C_{60}$ the fullerene role is as strong as in Xe@$C_{60}$, while in $Ce^{4+}$@$C_{60}$ and La@$C_{60}$ no structure appears.

## 4. Discussion and conclusion

We have predicted the existence of Endohedral Giant resonances. They result from strong fullerenes static and dynamic action upon atom, "caged" inside the fullerene. We have predicted strong interference effects in subvalent shell photoionization. It was shown that the fullerenes shell modifies impressively the angular anisotropy parameters. We demonstrated that depending upon the photoelectron's speed the atomic Giant resonance is either destroyed, as in Xe@$C_{60}$ or remains almost untouched, as in La@$C_{60}$.

We have considered above as a fullerene only $C_{60}$. As it was mentioned in the Introduction, it would be of interest to see the alteration of the photoionization cross-section if instead of $C_{60}$ other fullerenes, like $C_{70}$, $C_{76}$, $C_{82}$ or $C_{87}$ or fullerenes onions were considered. We do not know the shape and photoionization cross-sections of $C_{70}$, $C_{76}$, $C_{82}$ or $C_{87}$ and the position of the "caged" atoms inside the fullerenes. However, to have the feeling of the fullerenes shell effect upon photoionization of noble gas atoms, we can properly use the results for $C_{60}$ by scaling them to other radius, collectivized electrons number etc. Since the effects of radiative enhancement and oscillations due to reflection are sensitive to the radius, potential of the fullerene and number of electrons in it, one can expect considerable alteration of the presented above results.

It is essential to have in mind that while being caged, the atoms inside can be ionized. The electrons go to the fullerenes shell that became instead of a neutral, a negatively charged surface. This was not accounted for in our consideration of $Ce^{4+}$@$C_{60}$. The surface charge requires some modification in the accounting for the reflection of the photoelectron by the fullerenes shell.

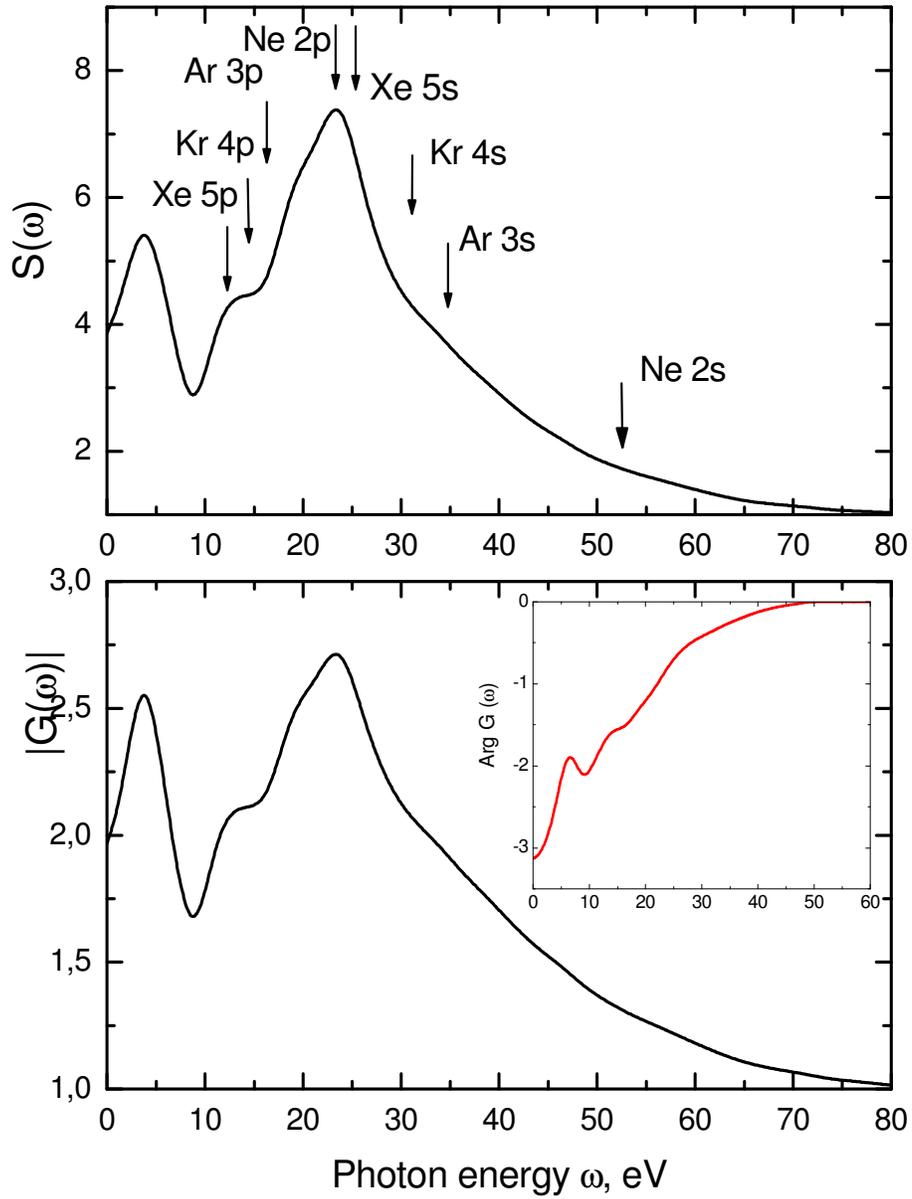

Fig.1. Radiation enhancement parameter $S(\omega)$, its amplitude's absolute value $\tilde{G}^d(\omega) \equiv |G(\omega)|$ and phase $\eta^d \equiv \arg G(\omega)$. Arrows denote the thresholds positions of corresponding outer $np$ and subvalent $ns$ subshells.

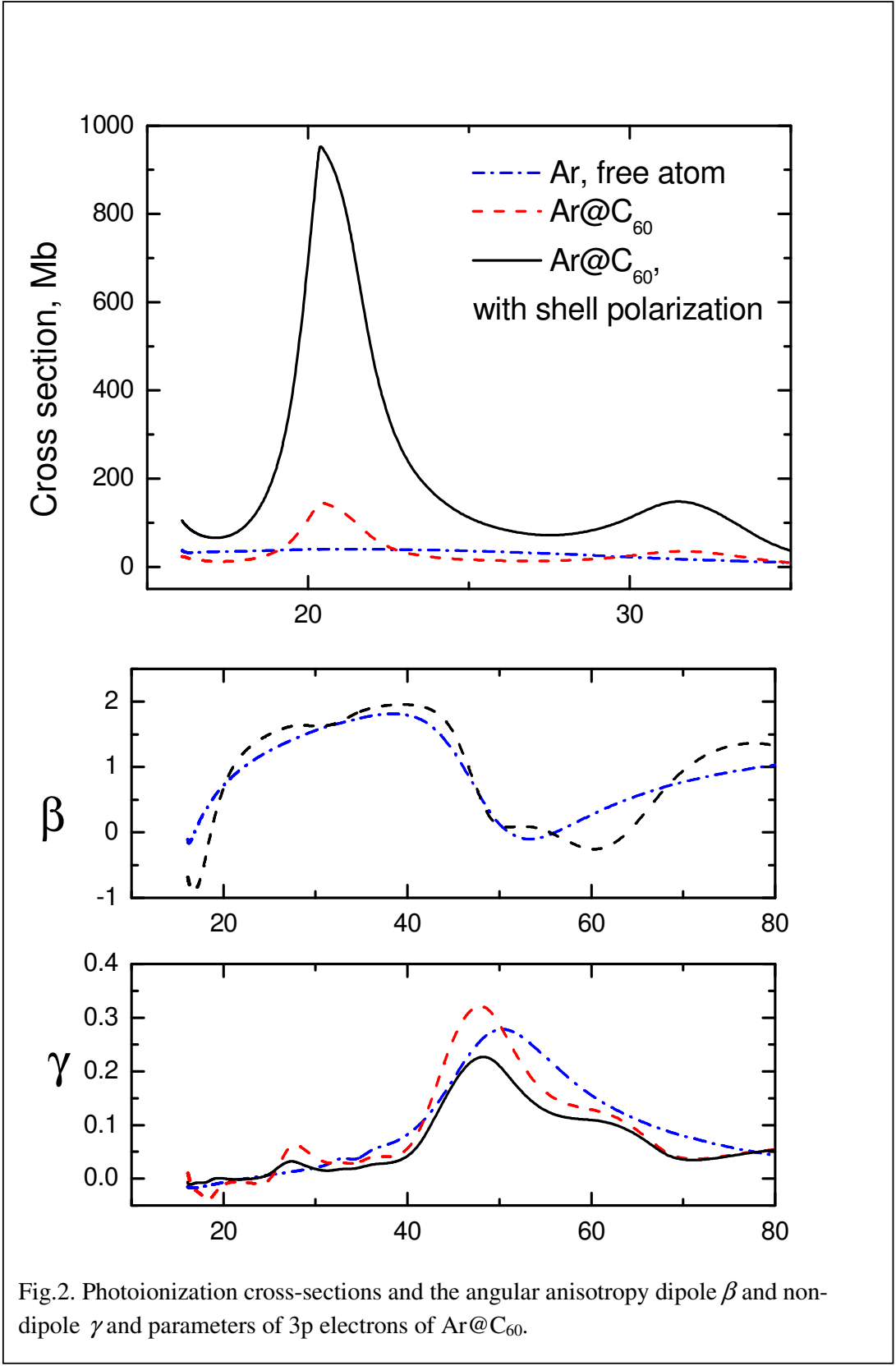

Fig.2. Photoionization cross-sections and the angular anisotropy dipole $\beta$ and non-dipole $\gamma$ and parameters of 3p electrons of Ar@$C_{60}$.

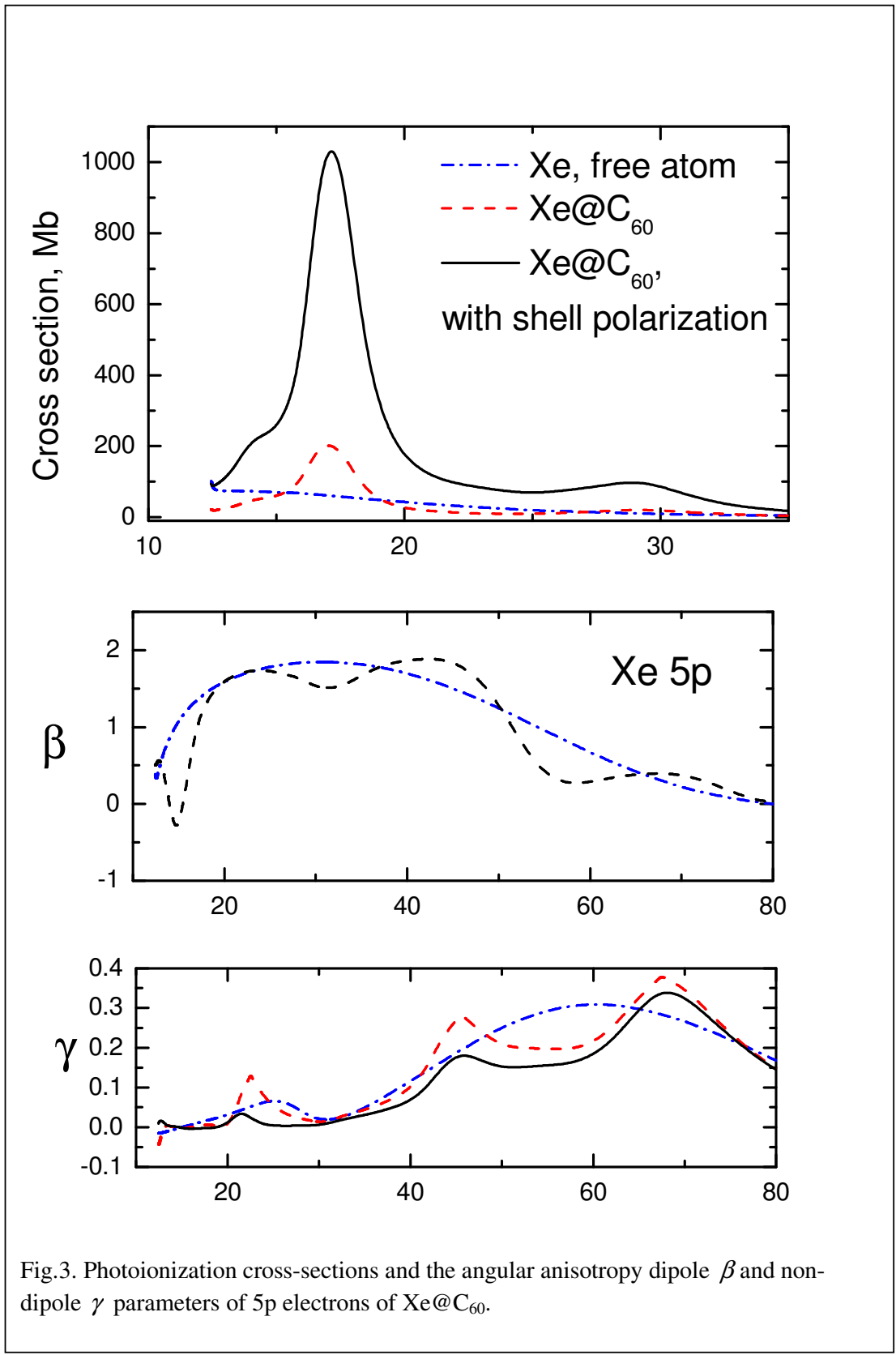

Fig.3. Photoionization cross-sections and the angular anisotropy dipole $\beta$ and non-dipole $\gamma$ parameters of 5p electrons of Xe@$C_{60}$.

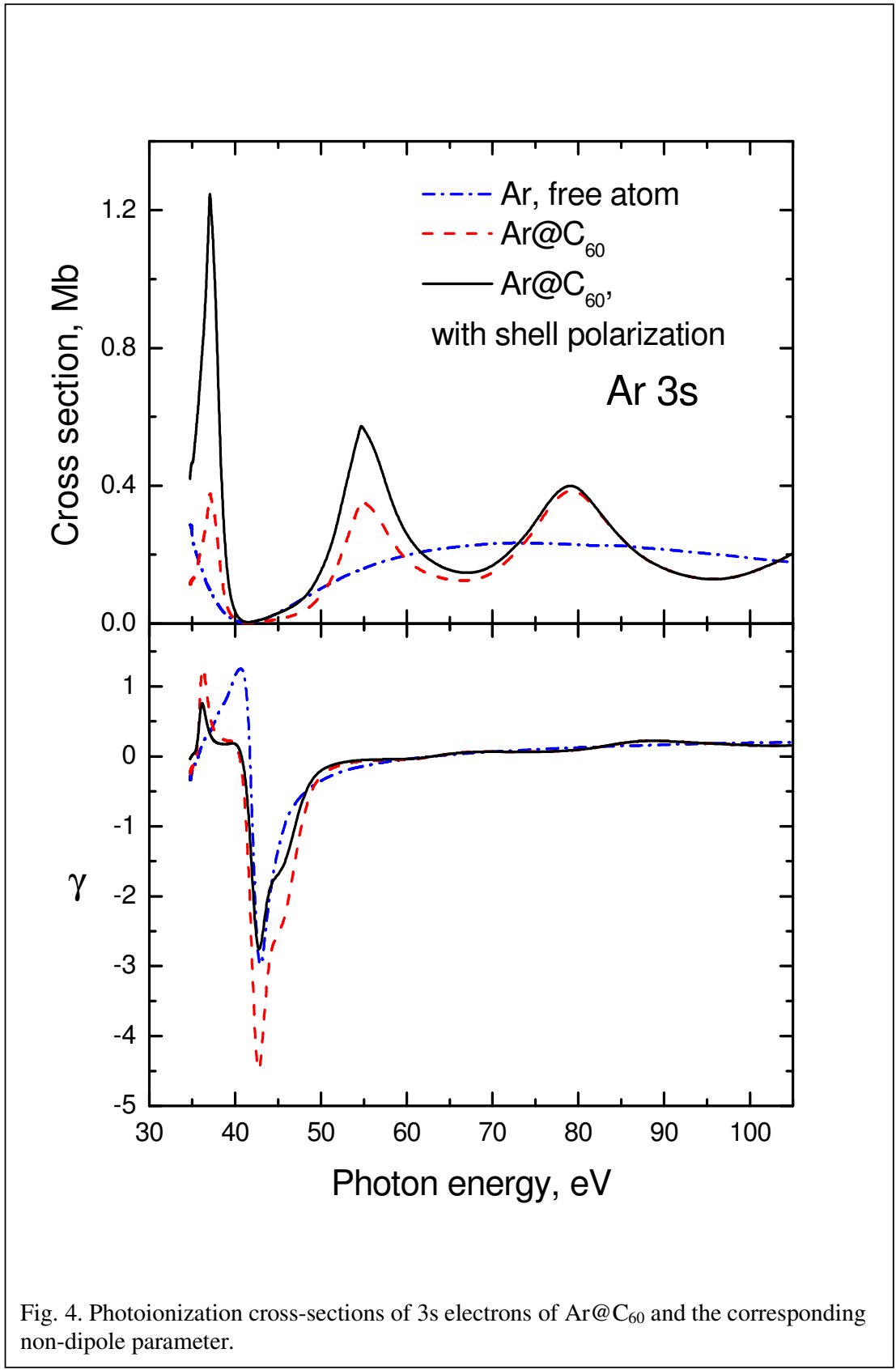

Fig. 4. Photoionization cross-sections of 3s electrons of Ar@C$_{60}$ and the corresponding non-dipole parameter.

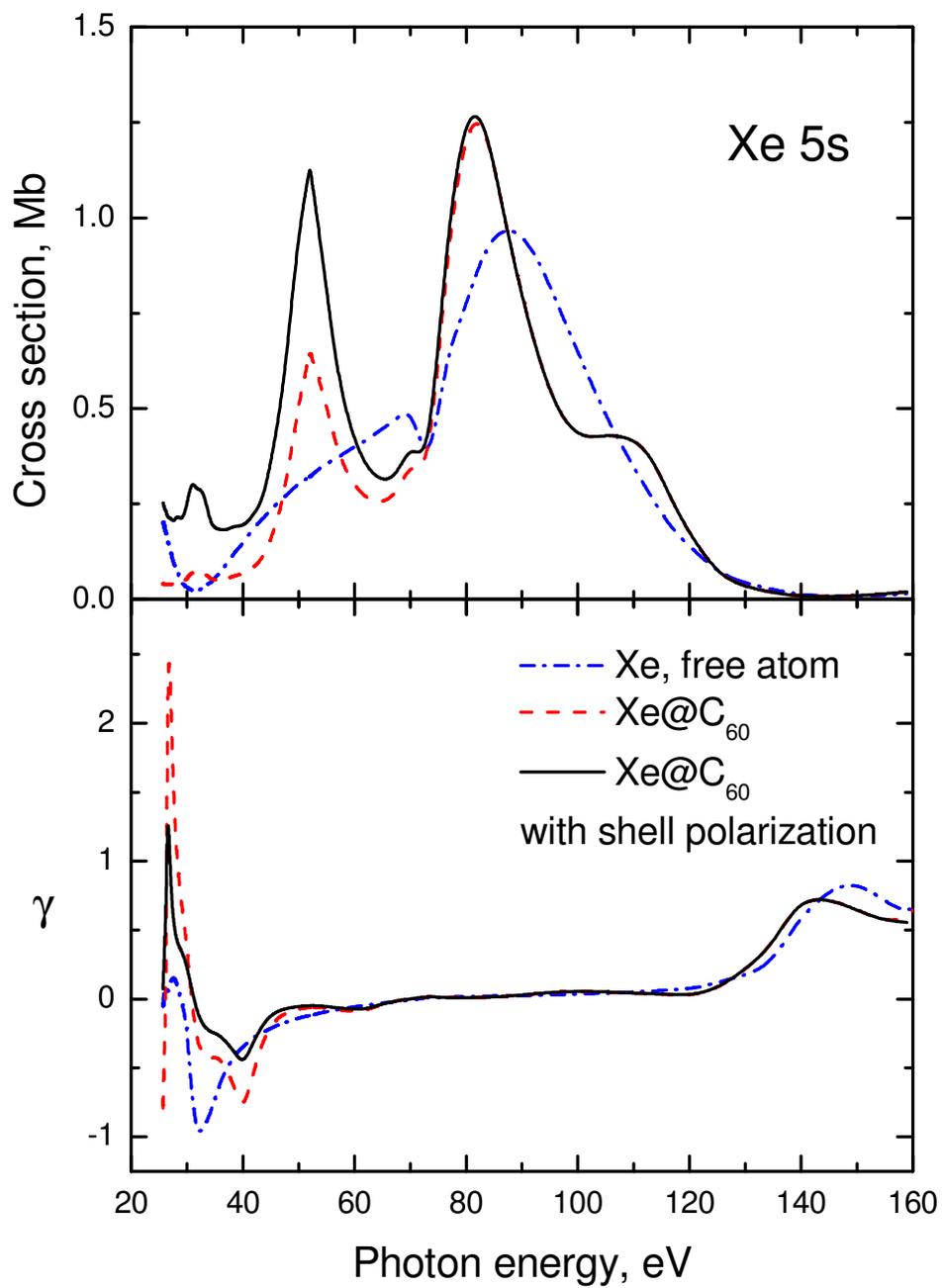

Fig. 5. Photoionization cross-sections of 5s electrons in Xe@$C_{60}$ and the corresponding non-dipole parameter.

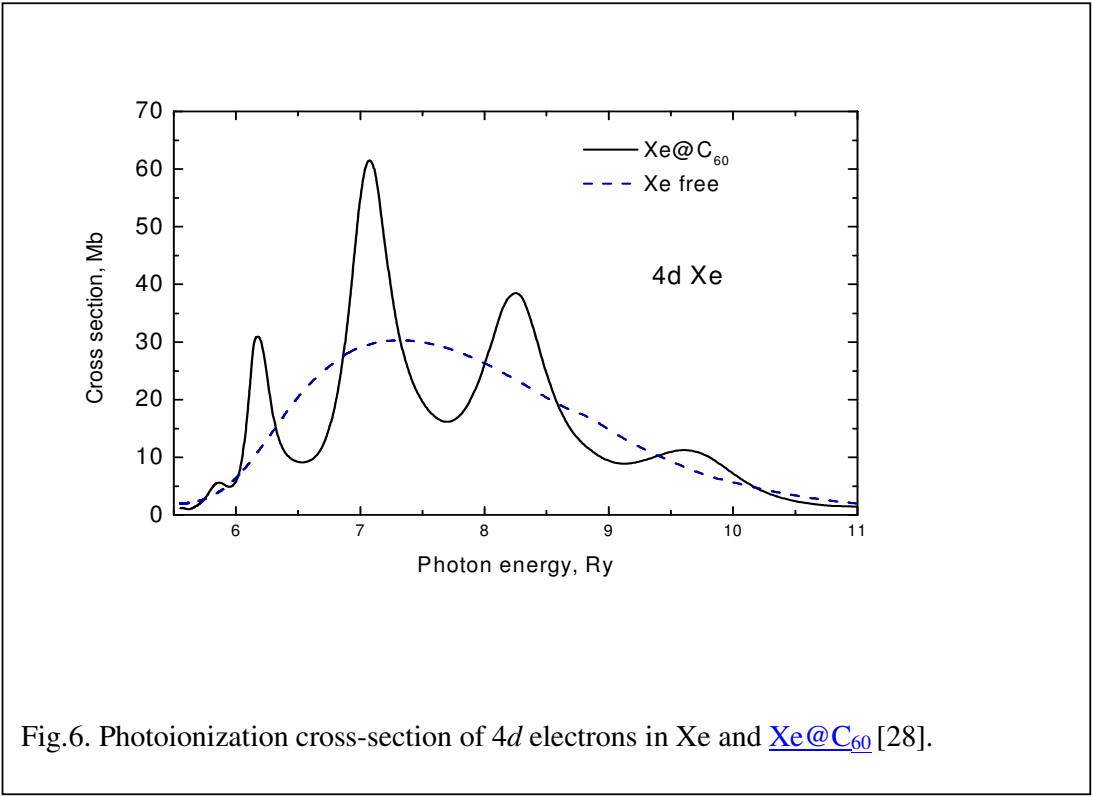

Fig.6. Photoionization cross-section of 4*d* electrons in Xe and Xe@C$_{60}$ [28].

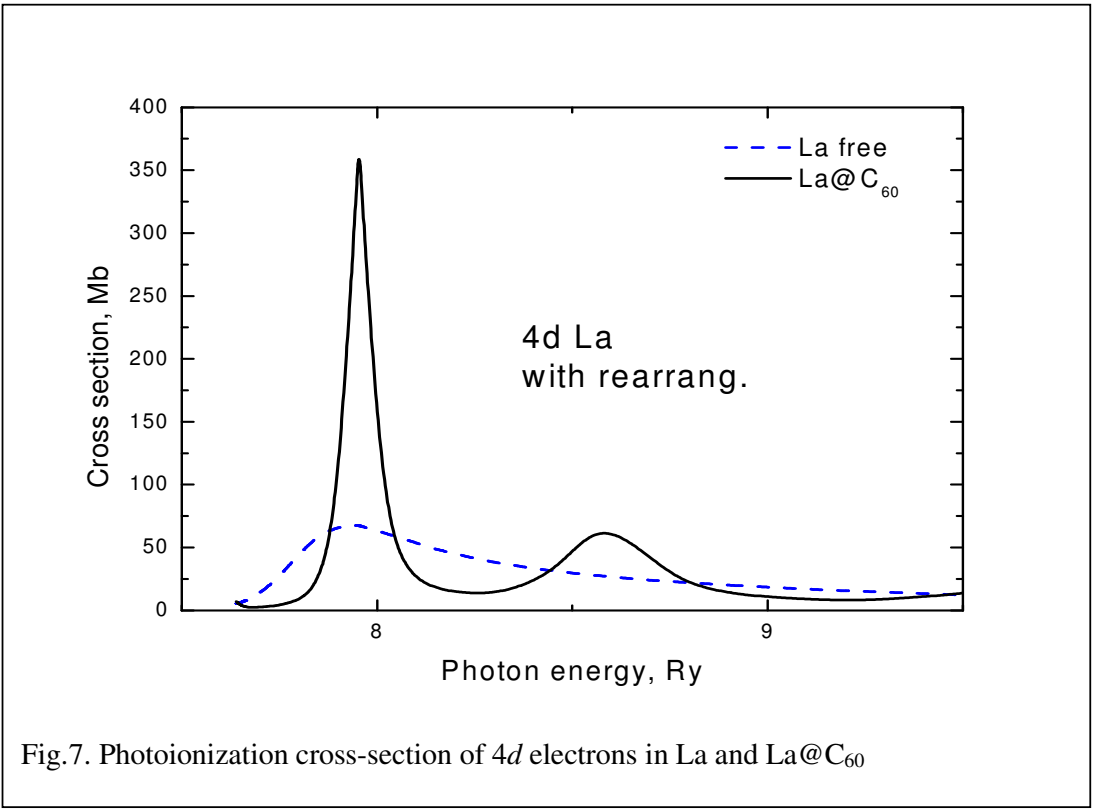

Fig.7. Photoionization cross-section of 4*d* electrons in La and La@C$_{60}$

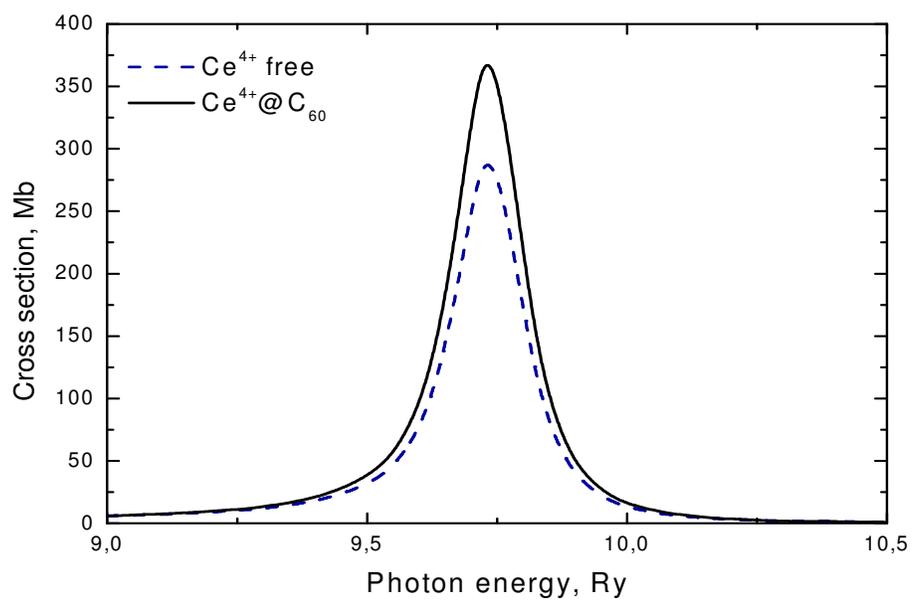

Fig.8. Photoionization cross-section of 4*d* electrons in $Ce^{4+}$ and $Ce^{4+}@C_{60}$.

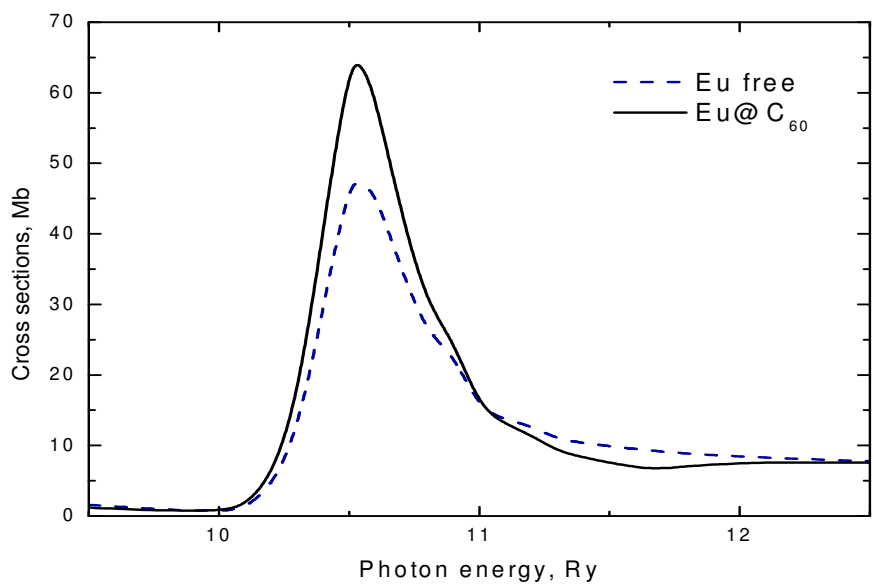

Fig.9. Photoionization cross-section of 4*d* electrons in Eu and $Eu@C_{60}$